\documentclass[12pt]{article}
\usepackage{epsf}

\newcommand{\bra}[1]{\left\langle{#1}\,\right|\,}
\newcommand{\ket}[1]{\vert\,{#1}\rangle}
\renewcommand{\bar}{\overline}
\newcommand{\ie}{{\it i.e.}}
\newcommand{\eg}{{\it e.g.}}

\newcommand{\half}  {\frac{1}{2}}

\def\ru1{\rule[-0.4truecm]{0mm}{1truecm}}

\begin {document}
\begin{flushright}
{\small
SLAC--PUB--8626\\
September 2000\\}
\end{flushright}

\begin{center}
{{\bf\LARGE   
The Light-Cone Fock Representation in QCD}\footnote{Work supported by
Department of Energy contract  DE--AC03--76SF00515.}}

\bigskip
Stanley J. Brodsky\\
{\sl Stanford Linear Accelerator Center \\
Stanford University, Stanford, California 94309\\
sjbth@slac.stanford.edu}\\
\medskip
\end{center}

\vfill

\begin{center}
{\bf\large   
Abstract }
\end{center}

The light-cone Fock-state representation of QCD
encodes the properties of a hadrons in terms of frame-independent
wavefunctions, providing
a systematic framework for
evaluating structure functions, exclusive hadronic matrix elements,
including time-like heavy hadron decay amplitudes, form factors, and
deeply virtual Compton scattering.  A new
type of jet production reaction, ``self-resolving diffractive
interactions" can provide direct information on the light-cone
wavefunctions of hadrons in terms of their quark and gluon degrees of
freedom as well as the composition of nuclei in terms of their nucleon
and mesonic degrees of freedom.  The relation of the intrinsic sea to the
light-cone wavefunctions is discussed.
The decomposition of light-cone
spin is illustrated for the quantum fluctuations of
an electron.

\vfill

\begin{center} 
{\it Invited talk presented at  \\
HD2000---From Hadrons to Strings} \\
 {\it Max-Planck Haus, Heidelberg, Germany }\\
{\it June 12--17, 2000}\\
\end{center}

\vfill

\vfill\eject


\normalsize

\section{Introduction}

The light-cone wavefunctions $\{\psi_{n/H}(x_i,{\vec
k_{\perp i}},\lambda_i)\}$ are the interpolating
amplitudes between a hadron and its quark and gluon degrees
of freedom.  For example, the
proton eigenstate of the light-cone Hamiltonian in QCD satisfies:
$ H^{QCD}_{LC} \ket{\Psi_p} = M^2_p \ket{\Psi_p}$.
The projection of
the proton's eigensolution $\ket{\Psi_p}$ on the color-singlet
$B = 1$, $Q = 1$ eigenstates $\{\ket{n} \}$
of the free Hamiltonian $H^{QCD}_{LC}(g = 0)$ gives the
light-cone Fock expansion: \cite{BrodskyLepage}
\begin{eqnarray}
\left\vert \Psi_p; P^+, {\vec P_\perp}, \lambda \right> &=&
\sum_{n \ge 3,\lambda_i}  \int \Pi^{n}_{i=1}
{d^2k_{\perp i} dx_i \over \sqrt{x_i} 16 \pi^3}
16 \pi^3 \delta\left(1- \sum^n_j x_j\right) \delta^{(2)}
\left(\sum^n_\ell \vec k_{\perp \ell}\right) \nonumber\\[1ex]
&&\left\vert n;
x_i P^+, x_i {\vec P_\perp} + {\vec k_{\perp i}}, \lambda_i\right >
\psi_{n/p}(x_i,{\vec k_{\perp i}},\lambda_i) \  .
\end{eqnarray}
The coordinates of the light-cone Fock wavefunctions $\psi_{n/H}(x_i,{\vec
k_{\perp i}},\lambda_i)$ are the light-cone
momentum fractions
$x_i = k^+_i/P^+$ and the transverse
momenta ${\vec k_{\perp i}}$ of its constituents.
The $\lambda_i$ label the light-cone spin $S^z$ projections of
the quarks and
gluons along the $z$ direction.  The physical gluon
polarization vectors
$\epsilon^\mu(k,\ \lambda = \pm 1)$ are specified in light-cone
gauge $A^+ =0$ by the conditions $k \cdot \epsilon = 0,\ \eta \cdot
\epsilon = \epsilon^+ = 0.$
Each light-cone Fock wavefunction satisfies conservation of the
$z$ projection of angular momentum:
$
J^z = \sum^n_{i=1} S^z_i + \sum^{n-1}_{j=1} l^z_j \ .
$
The sum over $S^z_i$
represents the contribution of the intrinsic spins of the $n$ Fock state
constituents.  The sum over orbital angular momenta
$l^z_j = -{\mathrm i} (k^1_j\frac{\partial}{\partial k^2_j}
-k^2_j\frac{\partial}{\partial k^1_j})$ derives from
the $n-1$ relative momenta.  This definition automatically excludes the
contribution to the orbital angular momentum due to the motion of the
center of mass, which is not an intrinsic property of the
hadron.\cite{Brodsky:2000ii}
A comprehensive review of light-cone quantization and physics 
can be found  in the review.\cite{bpp97}

In the light-cone Hamiltonian (front form) method, a quantum field theory
is quantized at a fixed light-cone time $\tau = t +z/c$.
\cite{Dirac:1949cp} The generator of light-cone time translations $P^-
=P^0 - P_z = i {\partial
\over
\partial
\tau}$ defines the light cone Hamiltonian of QCD.  The light-cone momenta
$P^+ = P^0 + P^z$ and $P_\perp$ are
kinematical and commute with $P^-$.
It is very useful to define the
invariant operator
$ H_{LC} = P^+ P^- - {\vec P_\perp}^2$ since its set of
eigenvalues
${\mathcal M}^2_n$ enumerates the
bound state and continuum (scattering state) mass spectrum.
\eject

Gauge theories are usually quantized in light-cone gauge $A^+ = 0$ in
which the propagating gauge fields have transverse polarization.  An
additional advantage is the absence of ghost fields, even in the case of
non-Abelian theories.  Srivastava and I \cite{Srivastava:2000gi} have
recently used the Dirac procedure for canonically quantizing the theory
and shown that light-cone gauge automatically incorporates the Lorentz
gauge condition
$(\partial\cdot A)=0$.  The interactions in the light-cone Hamiltonian of
gauge theories contain additional light-cone-time instantaneous fermion
and gauge field interactions, analogous to Coulomb interactions in the
instant form.  However, one can also use the Dyson-Wick method to derive
the Feynman covariant rules for gauge theory in light-cone gauge with
causal propagators and no ghost fields.  \cite{Srivastava:2000gi}
We have also shown that one can also effectively quantize QCD on the
light-cone in the covariant Feynman gauge.\cite{Srivastava:2000gi}

A crucial advantage of the light-cone formalism is the
Lorentz frame-indepen\-dence of the light-cone wavefunctions.  Knowing the
wavefunction for one momentum $P^\mu$ is sufficient to determining the
wavefunction for any other momentum $P^{\mu \prime}$.  Note that the
variables $x_i$ and $\vec k_{\perp i}$ are relative coordinates,
independent of the hadron's momentum $P^+, P_\perp.$ The actual momenta
of the constituents which appear in the $n-$particle Fock states are
$p^+_i = x_i P^+$ and ${\vec p_{\perp i}} = x_i {\vec P_\perp} + {\vec
k_{\perp i}}.$ In light-cone quantization all $k^+_i$ are positive.
Since
the sum of plus momentum is conserved by the local interactions, vacuum
fluctuations cannot occur.  Thus the physical vacuum in light-cone
quantization coincides with the perturbative vacuum, no contributions to
matrix elements from vacuum fluctuations occur.

The frame-independence of the light-cone wavefunctions
can be contrasted with the equal-time wavefunctions obtained in
the conventional equal-time quantization.  In the ``instant form",  Lorentz
boosts are dynamical---boosting the hadron from one frame to another
requires a new solution of the non-perturbative Hamiltonian equation.
Light-cone quantization thus provides a frame-independent,
relativistic quantum-mechanical description of hadrons which encodes
multi-quark and gluon momentum, helicity, and flavor correlations in the
form of universal process-independent hadron wavefunctions.

Hadronic amplitudes can be generally computed in the light-cone
Hamiltonian formalism from the convolution of the light-cone
wavefunctions with the underlying irreducible quark-gluon amplitudes.
For example, spacelike form factors and other hadronic matrix
elements of local operators can
be expressed as simple overlaps of the initial and final state light-cone
wavefunctions with the same number $n = n^\prime$ of Fock constituents.
Particle creation or absorption into the vacuum is not allowed because of
the positivity of the $k^+$.  Diagrams in which the current creates or
annihilates pairs is forbidden in the special frame
$q^+ = 0, q_\perp^2 = Q^2 = -q^2$
\cite{Drell:1970km,West:1970av} for space-like momentum transfer.  One
can also take matrix elements of ``plus" components of currents such as
$J^+$ and
$T^{++}$ to avoid instantaneous contributions to the current
operator.

The simplicity of the diagonal representation of light cone matrix
elements such as $\bra{ p+q } J^+ \ket{ p} $ in the front form is in
striking contrast with the equal-time formalism.  In the instant form the
evaluation of current matrix elements not only requires overlaps of
states with different particle number, but it also requires the
computation of time-ordered amplitudes in which the photon
interacts with charged particles arising from vacuum fluctuations,
{\em e.g.},
$ \ket {0} \to \ket {q(\vec p_a)  \bar q(\vec p_b) g(\vec p_g)}$
with $\vec p_g  + \vec p_a + \vec p_b =
\vec 0$, particles which then are absorbed by the final-state equal-time
wavefunctions.  Thus in the instant form, current matrix elements are not
determined by the equal-time Fock wavefunctions alone.

Hwang, Ma, Schmidt and I \cite{Brodsky:2000ii} have recently used the
light-cone wavefunction representation of gravitational form factors to
prove that the anomalous moment coupling
$B(0)$ to a graviton vanishes for any composite system.
This remarkable result was first derived from the
equivalence principle by Okun and Kobzarev \cite{Okun};  see also
\cite{Ji:1996kb,Ji:1997ek,Ji:1997nm,Teryaev:1999su}.
In our proof we show that -- after summing over the couplings of the
gravitons to each of the $n$ constituents -- the contribution to
$B(0)$ vanishes identically for each Fock component $\psi_n$ due to of
the Lorentz boost properties of the light-cone Fock
representation.\cite{Brodsky:2000ii}

Exclusive
semi-leptonic
$B$-decay amplitudes involving timelike currents such as $B\rightarrow A
\ell
\bar{\nu}$ can also be evaluated exactly
in the light-cone Fock representation.\cite{Brodsky:1999hn,Ji:1999gt} In
this case, the timelike decay matrix elements require the computation of
both the diagonal matrix element $n \rightarrow n$ where parton number is
conserved and the off-diagonal $n+1\rightarrow n-1$ convolution such that
the current operator annihilates a $q{\bar{q'}}$ pair in the initial $B$
wavefunction.  This term is a consequence of the fact that the time-like
decay $q^2 = (p_\ell + p_{\bar{\nu}} )^2 > 0$ requires a positive
light-cone momentum fraction $q^+ > 0$.  As I shall discuss below, one can
also give a light-cone wavefunction representation of the deeply virtual
Compton amplitude.\cite{DVCSHwang}
\eject

Given the light-cone wavefunctions, one can compute the
moments of the helicity and
transversity distributions measurable in polarized deep inelastic
experiments.\cite{Lepage:1980fj} For example,
the polarized quark distributions at resolution $\Lambda$ correspond to
\begin{eqnarray}
q_{\lambda_q/\Lambda_p}(x, \Lambda)
&=& \sum_{n,q_a}
\int\prod^n_{j=1}{ dx_j d^2 k_{\perp j}\over 16 \pi^3} \sum_{\lambda_i}
\vert \psi^{(\Lambda)}_{n/H}(x_i,\vec k_{\perp i},\lambda_i)\vert^2
\nonumber\\
&& \times 16 \pi^3 \delta\left(1- \sum^n_i x_i\right) \delta^{(2)}
\left(\sum^n_i \vec k_{\perp i}\right)\\
&& \times
\delta(x - x_q) \delta_{\lambda_a, \lambda_q}
\Theta(\Lambda^2 - {\mathcal M}^2_n)\ , \nonumber
\end{eqnarray}
where the sum is over all quarks $q_a$ which match the quantum
numbers, light-cone momentum fraction $x,$ and helicity of the struck
quark.  Similarly, the distribution of spectator
particles in the final state which could be measured in the proton
fragmentation region in deep inelastic scattering at an electron-proton
collider are in principle encoded in the light-cone wavefunctions.
More generally, all multi-quark and gluon
correlations in the bound state are represented by the light-cone
wavefunctions.  Thus in principle, all of the complexity of a hadron is
encoded in the light-cone Fock representation, and the light-cone Fock
representation is thus a representation of the underlying quantum field
theory.

The key non-perturbative input for exclusive
processes at high momentum transfer is the gauge and frame independent
hadron distribution amplitude \cite{Lepage:1979zb,Lepage:1980fj} defined
as the integral of the valence (lowest particle number) Fock wavefunction;
\eg\ for the pion
\begin{equation}
\phi_\pi (x_i,\Lambda) \equiv \int d^2k_\perp\, \psi^{(\Lambda)}_{q\bar
q/\pi} (x_i, \vec k_{\perp i},\lambda)
\label{eq:f1}
\end{equation}
where the global cutoff $\Lambda$ in invariant mass is identified with the
resolution
$Q$. The distribution amplitude controls leading-twist exclusive
amplitudes at high momentum transfer, and it can be related to the
gauge-invariant Bethe-Salpeter wavefunction at equal light-cone time.  The
logarithmic evolution of hadron distribution amplitudes
$\phi_H (x_i,Q)$ can be derived from the perturbatively-computable tail
of the valence light-cone wavefunction in the high transverse momentum
regime.\cite{Lepage:1979zb,Lepage:1980fj} The conformal basis for the
evolution of the three-quark distribution amplitudes
for the baryons~\cite{Lepage:1979za} in terms of conformal eigensolutions
of the evolution kernel, the Jacobi polynomials, has recently been
obtained by V. Braun {\em et al.}\cite{Braun:1999te}
The asymptotic solution for the proton resembles a scalar $I=
0$ diquark structure.

\section{A perturbative example}

Recently Hwang, Ma,  Schmidt, and I have shown that
the light-cone wavefunctions generated by the radiative corrections to
the electron in QED provides an ideal system for understanding
the spin and angular momentum decomposition of relativistic
systems.\cite{Brodsky:2000ii} The model
is patterned after the quantum structure which
occurs in the one-loop Schwinger
${\alpha / 2 \pi} $ correction to the electron magnetic
moment.\cite{Brodsky:1980zm} In effect, we can represent a spin-$\half$ ~
system as a composite of a spin-$\half$ ~ fermion and spin-one vector boson
with arbitrary masses.  A similar model has recently been used to
illustrate the matrix elements and evolution of light-cone helicity and
orbital angular momentum operators.\cite{Harindranath:1999ve} This
representation of a composite system is particularly useful because it
is based on two constituents but yet is totally relativistic.  We can
also explicitly compute the form factors
$F_1(q^2)$ and $F_2(q^2)$ of the electromagnetic current, and the
various contributions to the form factors
$A(q^2)$ and $B(q^2)$ of the energy-momentum tensor.

For example, the two-particle Fock state for an electron with $J^z =
+ {1\over 2}$ has four possible spin
combinations:~\cite{Lepage:1980fj,Brodsky:1980zm}
\begin{equation}
\left
\{ \begin{array}{l}
\psi^{\uparrow}_{+\frac{1}{2}\, +1} (x,{\vec k}_{\perp})=-{\sqrt{2}}
\frac{(-k^1+{\mathrm i} k^2)}{x(1-x)}\,
\varphi \ , [\ell^z = -1]\\
\psi^{\uparrow}_{+\frac{1}{2}\, -1} (x,{\vec k}_{\perp})=-{\sqrt{2}}
\frac{(+k^1+{\mathrm i} k^2)}{1-x }\,
\varphi \ ,[\ell^z = +1] \\
\psi^{\uparrow}_{-\frac{1}{2}\, +1} (x,{\vec k}_{\perp})=-{\sqrt{2}}
(M-{m\over x})\,
\varphi \ ,[\ell^z = 0] \\
\psi^{\uparrow}_{-\frac{1}{2}\, -1} (x,{\vec k}_{\perp})=0\ ,
\end{array}
\right.
\label{vsn2}
\end{equation}
where
\begin{equation}
\varphi=\varphi (x,{\vec k}_{\perp})=\frac{ e/\sqrt{1-x}}{M^2-({\vec
k}_{\perp}^2+m^2)/x-({\vec k}_{\perp}^2+\lambda^2)/(1-x)}\ .
\label{wfdenom}
\end{equation}
Each configuration satisfies
the spin sum rule: $J^z=S^z_{\rm f}+s^z_{\rm b} + l^z = +{1\over 2}$.
The sign of the helicity of the
electron is retained by the leading photon at $x_\gamma = 1- x \to 1$.
Note that in the non-relativistic limit, the transverse motion of the
constituents can be neglected, and we have only the
$\left|+\frac{1}{2}\right> \to
\left|-\frac{1}{2}\, +1\right>$ configuration which is the
non-relativistic quantum state for the spin-half system composed of
a fermion and a spin-1 boson constituents.  The fermion
constituent has spin projection in the opposite
direction to the spin $J^z$ of the whole system.
However, for ultra-relativistic binding in which
the transversal motions of the constituents are large compared to the
fermion masses,  the
$\left|+\frac{1}{2}\right> \to \left|+\frac{1}{2}\, +1\right>$
and
$\left|+\frac{1}{2}\right> \to \left|+\frac{1}{2}\, -1\right>$
configurations dominate
over the $\left|+\frac{1}{2}\right> \to \left|-\frac{1}{2}\, +1\right>$
configuration.  In this case
the fermion constituent has
spin projection parallel to $J^z$.

The spin
structure of perturbative theory provides a
template for the numerator structure of the light-cone wavefunctions
even for composite systems since the equations which couple
different Fock components mimic the perturbative form.
The structure of the electron's Fock state in perturbative QED shows that
it is natural to have a negative contribution from relative orbital
angular momentum which balances the $S_z$ of its photon constituents.
We can thus expect a large orbital contribution to the proton's
$J_z$ since gluons carry roughly half of the proton's
momentum, thus providing insight into the ``spin crisis" in QCD.

\section{Light-cone Representation of Deeply Virtual \hfil\break
Compton Scattering}

The virtual Compton scattering process ${d\sigma\over dt}(\gamma^*
p \to \gamma p)$ for large initial photon virtuality
$q^2=-Q^2$ has extraordinary sensitivity to
fundamental features of the proton's structure.  Even though the final
state photon is on-shell, the deeply virtual process probes the
elementary quark structure of the proton near the light cone as an
effective local current.  In contrast to deep inelastic scattering, which
measures only the absorptive part of the forward virtual Compton
amplitude, deeply virtual Compton scattering allows the measurement of
the phase and spin structure of proton matrix elements of the current
correlator for general momentum transfer squared $t$.

To leading order in $1/Q$, the deeply virtual Compton scattering
amplitude factorizes as the convolution in $x$ of the
amplitude $t^{\mu \nu}$ for hard Compton scattering on a quark line with
the generalized Compton form factors $f_i(x,t,\zeta)$ of the target
proton.\cite{Ji9697,Ji:1998pc,Radyushkin:1996nd,Ji:1998xh,%
Guichon:1998xv,Vanderhaeghen:1998uc,Radyushkin:1999es,%
Collins:1999be,Diehl98,Diehl:1999tr,Blumlein:2000cx}
Here
$x$ is the light-cone momentum fraction of the struck quark, and
$\zeta= Q^2/2 P\cdot q$ plays the role of the Bjorken variable.
Integrals of these quantities over $x$ are independent
of $\zeta$ and are equal to the gravitational form factors $A_q(t)$ and
$B_q(t)$ where the graviton couples only to quarks.

Recently, Markus Diehl, Dae Sung Hwang, and I \cite{DVCSHwang} have shown
how the deeply virtual Compton amplitude can be evaluated explicitly in
the Fock state representation using the matrix elements of the currents
and the boost properties of the light-cone wavefunctions.
We choose the frame where
$
P_I=(P^+\ ,\ {\vec P_\perp}\ ,\ P^-)\ =\ \left(\ P^+\ ,\ {\vec 0_\perp}\
,\ {M^2\over P^+}\ \right)\ ,$ and
$P_F=(P'^+\ ,\ {\vec P'_\perp}\ ,\ P'^-)\ =\
\left( (1-\zeta)P^+\ ,\ -{\vec \Delta_\perp}\ ,\ {(M^2+{\vec
\Delta_\perp}^2)\over (1-\zeta)P^+}\right)\ .$
The incident space-like
photon carries $q^+ = 0$ and $q^2= - Q^2 = - {\vec q_\perp}^2$ so that no
light-cone time-ordered amplitudes involving the splitting of the
incident photon can occur.

The diagonal (parton-number-conserving)
contribution to the generalized form factors for deeply virtual Compton
amplitude in the domain\cite{Diehl98,Diehl:1999tr,Muller:1994fv}
$\zeta\le x_1\le 1$ is:\\
\begin{eqnarray}
&&{\sqrt{1-\zeta}}f_{1\, (n\to n)}(x_1,t,\zeta)\,
-\, {\zeta^2\over 4{\sqrt{1-\zeta}}} f_{2\, (n\to n)}(x_1,t,\zeta)
\nonumber\\
 &=&
\sum_{n, ~ \lambda}
\prod_{i=1}^{n}
\int^1_0 dx_{i(i\ne 1)} \int {d^2{\vec{k}}_{\perp i} \over 2 (2\pi)^3 }
~ \delta\left(1-\sum_{j=1}^n x_j\right) ~ \delta^{(2)}
\left(\sum_{j=1}^n {\vec{k}}_{\perp j}\right)  \\[1ex]
&&\times
\psi^{\uparrow \ *}_{(n)}(x^\prime_i, {\vec{k}}^\prime_{\perp i},\lambda_i)
~ \psi^{\uparrow}_{(n)}(x_i, {\vec{k}}_{\perp i},\lambda_i)
(\sqrt{1-\zeta})^{1-n}, \nonumber
\label{t1}
\end{eqnarray}
\begin{eqnarray}
&&
{\sqrt{1-\zeta}}\,\left(\, 1+{\zeta\over 2(1-\zeta)}\,\right)\,
{(\Delta^1-{\mathrm i} \Delta^2)\over 2M}f_{2\, (n\to n)}(x_1,t,\zeta)
\nonumber\\
 &=&
\sum_{n, ~ \lambda}
\prod_{i=1}^{n}
\int^1_0 dx_{i(i\ne 1)} \int {d^2{\vec{k}}_{\perp i} \over 2 (2\pi)^3 }
~ \delta\left(1-\sum_{j=1}^n x_j\right) ~ \delta^{(2)}
\left(\sum_{j=1}^n {\vec{k}}_{\perp j}\right)  \\[1ex]
&&\qquad\qquad\qquad\times
\psi^{\uparrow \ *}_{(n)}(x^\prime_i, {\vec{k}}^\prime_{\perp i},\lambda_i)
~ \psi^{\downarrow}_{(n)}(x_i, {\vec{k}}_{\perp i},\lambda_i)
(\sqrt{1-\zeta})^{1-n} , \nonumber
\label{t1f2}
\end{eqnarray}
where
\begin{equation}
\left\{ \begin{array}{lll}
x^\prime_1 = {x_1-\zeta \over 1-\zeta}\, ,\
&{\vec{k}}^\prime_{\perp 1} ={\vec{k}}_{\perp 1}
- {1-x_1\over 1-\zeta} {\vec{\Delta}}_\perp
&\mbox{for the struck quark,}\\[1ex]
x^\prime_i = {x_i\over 1-\zeta}\, ,\
&{\vec{k}}^\prime_{\perp i} ={\vec{k}}_{\perp i}
+ {x_i\over 1-\zeta} {\vec{\Delta}}_\perp
&\mbox{for the $ (n-1)$ spectators.}
\end{array}\right.
\label{t2}
\end{equation}

One also has contributions from the $n+1 \to n-1$ off-diagonal Fock
transitions in the ``ERBL" domain $0\le x_1\le \zeta$:
\begin{eqnarray}
&&{\sqrt{1-\zeta}}f_{1\, (n+1\to n-1)}(x_1,t,\zeta)\,
-\, {\zeta^2\over 4{\sqrt{1-\zeta}}} f_{2\, (n+1\to n-1)}(x_1,t,\zeta)
\nonumber\\
 &=&
\sum_{n, ~ \lambda}
\int^1_0 dx_{n+1}
\int {d^2{\vec{k}}_{\perp 1} \over 2 (2\pi)^3 }
\int {d^2{\vec{k}}_{\perp n+1} \over 2 (2\pi)^3 }
\prod_{i=2}^{n}
\int^1_0 dx_{i} \int {d^2{\vec{k}}_{\perp i} \over 2 (2\pi)^3 }
\nonumber\\[2ex]
&&\times \delta\left(1-\sum_{j=1}^{n+1} x_j\right) ~
\delta^{(2)}\left(\sum_{j=1}^{n+1} {\vec{k}}_{\perp j}\right)
[\sqrt{1-\zeta}]^{1-n}
\\[2ex]
&&\times\psi^{\uparrow\ *}_{(n-1)}
(x^\prime_i,{\vec{k}}^\prime_{\perp i},\lambda_i)
~ \psi^{\uparrow}_{(n+1)}(\{x_1, x_i, x_{n+1} = \zeta - x_{1}\},
\nonumber\\[2ex]
&&\qquad\ \ \{ {\vec{k}}_{\perp 1},
{\vec{k}}_{\perp i},
{\vec{k}}_{\perp n+1} = {\vec{\Delta}}_\perp-{\vec{k}}_{\perp 1}\},
\{\lambda_1,\lambda_{i},\lambda_{n+1} = - \lambda_{1}\})
, \nonumber
\end{eqnarray}

\begin{eqnarray}
&&
{\sqrt{1-\zeta}}\,\Big(\, 1+{\zeta\over 2(1-\zeta)}\,\Big)\,
{(\Delta^1-{\mathrm i} \Delta^2)\over 2M}f_{2\, (n+1\to n-1)}(x_1,t,\zeta)
\nonumber\\
 &=&
\sum_{n, ~ \lambda}
\int^1_0 dx_{n+1}
\int {d^2{\vec{k}}_{\perp 1} \over 2 (2\pi)^3 }
\int {d^2{\vec{k}}_{\perp n+1} \over 2 (2\pi)^3 }
\prod_{i=2}^{n}
\int^1_0 dx_{i} \int {d^2{\vec{k}}_{\perp i} \over 2 (2\pi)^3 }
\nonumber\\[2ex]
&&\qquad\qquad\qquad\times \delta\left(1-\sum_{j=1}^{n+1} x_j\right) ~
\delta^{(2)}\left(\sum_{j=1}^{n+1} {\vec{k}}_{\perp j}\right)
[\sqrt{1-\zeta}]^{2-n}
\\[2ex]
&&\qquad\qquad\qquad\times\psi^{\uparrow\ *}_{(n-1)}
(x^\prime_i,{\vec{k}}^\prime_{\perp i},\lambda_i)
~ \psi^{\downarrow}_{(n+1)}(\{x_1, x_i, x_{n+1} = \zeta - x_{1}\},
\nonumber\\[2ex]
&&\qquad\qquad\ \ \{ {\vec{k}}_{\perp 1},
{\vec{k}}_{\perp i},
{\vec{k}}_{\perp n+1} = {\vec{\Delta}}_\perp-{\vec{k}}_{\perp 1}\},
\{\lambda_1,\lambda_{i},\lambda_{n+1} = - \lambda_{1}\})
, \nonumber
\end{eqnarray}
where $i=2,3,\cdots ,n$
label the $n-1$ spectator
partons which appear in the final-state hadron wavefunction
with
\begin{equation}
x^\prime_i = {x_i\over 1-\zeta}\, ,\qquad
{\vec{k}}^\prime_{\perp i} ={\vec{k}}_{\perp i}
+ {x_i\over 1-\zeta} {\vec{\Delta}}_\perp \ .
\end{equation}

The above representation is the general form for the generalized form
factors of the deeply virtual Compton amplitude for any composite system.
Thus given the light-cone Fock state wavefunctions of the eigensolutions
of the light-cone Hamiltonian, we can compute the amplitude for virtual
Compton scattering including all spin correlations.  The formulae are
accurate to leading order in $1/Q^2$.  Radiative corrections to the quark
Compton amplitude of order
$\alpha_s(Q^2)$ from diagrams in which a hard gluon interacts between
the two photons have also been neglected.

\section {The Physics of Non-Valence Fock states}

The light-cone representation allows one to understand the role of
higher particle number states in hadron phenomenology.
One can identify two contributions to the heavy quark sea, the
``extrinsic'' contributions which correspond to ordinary gluon splitting,
and the ``intrinsic" sea which is multi-connected via gluons to the
valence quarks.  The leading
$1/m_Q^2$ contributions to the intrinsic sea of the proton in the heavy
quark expansion are proton matrix elements of the
operator~\cite{Franz:2000ee}
$\eta^\mu \eta^\nu G_{\alpha \mu} G_{\beta \nu} G^{\alpha \beta}$ which
in light-cone gauge $\eta^\mu A_\mu= A^+= 0$ corresponds to three or four
gluon exchange between the heavy-quark loop and the proton constituents
in the forward virtual Compton amplitude.  The intrinsic sea is thus
sensitive to the hadronic bound-state structure.\cite{Brodsky:1981se} The
maximal contribution of the intrinsic heavy quark occurs at $x_Q \simeq
{m_{\perp Q}/ \sum_i m_\perp}$ where
$m_\perp = \sqrt{m^2+k^2_\perp}$;
\ie\ at large $x_Q$, since this minimizes the invariant mass ${\mathcal
M}^2_n$.
The
measurements of the charm structure function by the EMC experiment are
consistent with intrinsic charm at large $x$ in the nucleon with a
probability of order $0.6 \pm 0.3 \% $.\cite{Harris:1996jx} which is
consistent with recent estimates based on instanton
fluctuations.\cite{Franz:2000ee} Similarly, one can distinguish
intrinsic gluons which are associated with multi-quark interactions and
extrinsic gluon contributions associated with quark
substructure.\cite{Brodsky:1990db} One can also use this framework to
isolate the physics of the anomaly contribution to the Ellis-Jaffe sum
rule.\cite{Bass:1998rn} Neither gluons nor sea quarks are solely
generated by DGLAP evolution, and one cannot define a resolution scale
$Q_0$ where the sea or gluon degrees of freedom can be neglected.

The presence of intrinsic sea can have important consequences in $B$
decay, since the sea quarks can enter directly into the weak interaction
subprocess, thus effectively evading the hierarchy of the CKM
matrix.\cite{Gardner} Similarly, in a semi-exclusive reaction one can have
partial annihilation of a quark and anti-quark by the weak interaction
leaving the remaining partons to form the final hadrons.
\cite{Brodsky:1999hn}

It is usually assumed that a heavy quarkonium state such as the
$J/\psi$ always decays to light hadrons via the annihilation of its heavy quark
constituents to gluons.  However, as Karliner and I \cite{Brodsky:1997fj}
have shown, the transition $J/\psi \to \rho
\pi$ can also occur by the rearrangement of the $c \bar c$ from the $J/\psi$
into the $\ket{ q \bar q c \bar c}$ intrinsic charm Fock state of the $\rho$ or
$\pi$.  On the other hand, the overlap rearrangement integral in the
decay $\psi^\prime \to \rho \pi$ will be suppressed since the intrinsic
charm Fock state radial wavefunction of the light hadrons will evidently
not have nodes in its radial wavefunction.  This observation provides
a natural explanation of the long-standing puzzle~\cite{Brodsky:1987bb}
why the $J/\psi$ decays prominently to two-body pseudoscalar-vector final
states, breaking hadron helicity
conservation,~\cite{Brodsky:1981kj} whereas the
$\psi^\prime$ does not.

\section{Applications of Light-Cone Factorization to Hard QCD
\hfill\break Processes}

The light-cone
formalism provides a physical factorization scheme which conveniently
separates and factorizes soft non-perturbative physics from hard
perturbative dynamics in both exclusive and
inclusive reactions.\cite{Lepage:1980fj,Lepage:1979zb} In hard inclusive
reactions all intermediate states are divided according to ${\mathcal M}^2_n <
\Lambda^2 $ and
${\mathcal M}^2_n >
\Lambda^2 $ domains.  The lower mass regime is associated with the quark
and gluon distributions defined from the absolute squares of the LC
wavefunctions in the light cone factorization scheme.  In the high
invariant mass regime, intrinsic transverse momenta can be ignored, so
that the structure of the process at leading power has the form of hard
scattering on collinear quark and gluon constituents, as in the parton
model.  The attachment of gluons from the LC wavefunction to a propagator
in a hard subprocess is power-law suppressed in LC gauge, so that the
minimal quark-gluon particle-number subprocesses dominate.

There are many applications of this formalism:
{\it Exclusive Processes and Heavy Hadron Decays.} At high transverse
momentum an exclusive amplitudes factorize into a convolution of a hard
quark-gluon subprocess amplitudes $T_H$ with the hadron distribution
amplitudes
$\phi(x_i,\Lambda)$.  {\it Color Transparency.} Each Fock state interacts
distinctly; \eg\ Fock states with small particle number and small impact
separation have small color dipole moments and can traverse a nucleus
with minimal interactions.  This is the basis for the predictions for
color transparency \cite{BM} in hard quasi-exclusive reactions.  {\it
Diffractive vector meson photoproduction.} The light-cone Fock
wavefunction representation of hadronic amplitudes allows a simple
eikonal analysis of diffractive high energy processes, such as
$\gamma^*(Q^2) p \to \rho p$, in terms of the virtual photon and the vector
meson Fock state light-cone wavefunctions convoluted with the $g p \to g p$
near-forward matrix element.\cite{Brodsky:1994kf}
{\it Regge behavior of structure functions.} The light-cone wavefunctions
$\psi_{n/H}$ of a hadron are not independent of each other, but rather are
coupled via the QCD equations of motion.  The constraint of finite
``mechanical'' kinetic energy implies ``ladder relations"
which interrelate the light-cone wavefunctions of states differing by one
or two gluons.\cite{ABD} This in turn implies BFKL
Regge behavior of the polarized and unpolarized structure functions at $x
\rightarrow 0$.\cite{Mueller}
{\it Structure functions at large $x_{bj}$.} The behavior of structure functions
at $x \rightarrow 1$ is a highly off-shell light-cone
wavefunction configuration leading to quark-counting and
helicity-retention rules for the power-law behavior of the polarized and
unpolarized quark and gluon distributions in the endpoint domain.  The
effective starting point for the PQCD evolution of the structure
functions increases as $x \to 1$.  Thus evolution
is quenched at $ x \to 1$.\cite{Lepage:1980fj,BrodskyLepage,Dmuller}
{\it Hidden Color.}
The deuteron form factor at high $Q^2$ is sensitive to wavefunction
configurations where all six quarks overlap within an impact
separation $b_{\perp i} < {\mathcal O} (1/Q).$ The dominant
color configuration at large distances corresponds to the usual
proton-neutron bound state.  However, at small impact space
separation, all five Fock color-singlet components eventually
acquire equal weight, \ie, the deuteron wavefunction evolves to
80\%\ ``hidden color.'' The relatively large normalization of the
deuteron form factor observed at large $Q^2$ hints at sizable
hidden-color contributions.\cite{Farrar:1991qi} Hidden color components
can also play a predominant role in the reaction $\gamma d \to J/\psi p n$
at threshold if it is dominated by the multi-fusion process $\gamma g g
\to J/\psi$.

\section{Self-Resolved Diffractive Reactions}

Diffractive multi-jet dissociation of high energy hadrons in heavy
nuclei provides a novel way to measure the shape of the LC Fock
state wavefunctions and test color transparency.
For example, consider the
reaction \cite{Bertsch,MillerFrankfurtStrikman,Frankfurt:1999tq}
$\pi A \rightarrow {\rm Jet}_1 + {\rm Jet}_2 + A^\prime$
at high energy where the nucleus $A^\prime$ is left intact in its ground
state.  The transverse momenta of the jets balance so that
$
\vec k_{\perp i} + \vec k_{\perp 2} = \vec q_\perp < {R^{-1}}_A \ .
$
The light-cone longitudinal momentum fractions also need to add to
$x_1+x_2 \sim 1$ so that $\Delta p_L < R^{-1}_A$.  The process can
then occur coherently in the nucleus.  Because of color transparency, the
valence wavefunction of the pion with small impact separation, will
penetrate the nucleus with minimal interactions, diffracting into jet
pairs.\cite{Bertsch} The $x_1=x$, $x_2=1-x$ dependence of
the di-jet distributions will thus reflect the shape of the pion
valence light-cone wavefunction in $x$; similarly, the
$\vec k_{\perp 1}- \vec k_{\perp 2}$ relative transverse momenta of the
jets gives key information on the derivative of the underlying shape
of the valence pion
wavefunction.\cite{MillerFrankfurtStrikman,Frankfurt:1999tq,BHDP} The
diffractive nuclear amplitude extrapolated to
$t = 0$ should be linear in nuclear number $A$ if color transparency is
correct.  The integrated diffractive rate should then scale as $A^2/R^2_A
\sim A^{4/3}$.  Preliminary results on a diffractive dissociation
experiment of this type E791 at Fermilab using 500 GeV incident pions on
nuclear targets.\cite{Ashery:1999nq} appear to be consistent with color
transparency.\cite{Ashery:1999nq} The momentum fraction distribution of
the jets is consistent with a valence light-cone wavefunction of the pion
consistent with the shape of the asymptotic distribution amplitude,
$\phi^{\rm asympt}_\pi (x) =
\sqrt 3 f_\pi x(1-x)$.  Data from
CLEO \cite{Gronberg:1998fj} for the
$\gamma
\gamma^* \rightarrow \pi^0$ transition form factor also favor a form for
the pion distribution amplitude close to the asymptotic solution
\cite{Lepage:1979zb,Lepage:1980fj} to the perturbative QCD evolution
equation.

The diffractive dissociation of a hadron or nucleus can also occur via
the Coulomb dissociation of a beam particle on an electron beam (\eg\ at
HERA or eRHIC) or on the strong Coulomb field of a heavy nucleus (\eg\
at RHIC or nuclear collisions at the LHC).\cite{BHDP} The amplitude for
Coulomb exchange at small momentum transfer is proportional to the first
derivative $\sum_i e_i {\partial \over \vec k_{T i}} \psi$ of the
light-cone wavefunction, summed over the charged constituents.  The Coulomb
exchange reactions fall off less fast at high transverse momentum compared
to pomeron exchange reactions since the light-cone wavefunction is
effective differentiated twice in two-gluon exchange reactions.

It will also be interesting to study diffractive tri-jet production
using proton beams
$ p A \rightarrow {\rm Jet}_1 + {\rm Jet}_2 + {\rm Jet}_3 + A^\prime $ to
determine the fundamental shape of the 3-quark structure of the valence
light-cone wavefunction of the nucleon at small transverse
separation.\cite{MillerFrankfurtStrikman}
For example, consider the Coulomb dissociation of a high energy proton at
HERA.  The proton can dissociate into three jets corresponding to the
three-quark structure of the valence light-cone wavefunction.  We can
demand that the produced hadrons all fall outside an opening angle $\theta$
in the proton's fragmentation region.
Effectively all of the light-cone momentum
$\sum_j x_j \simeq 1$ of the proton's fragments will thus be
produced outside an ``exclusion cone".  This
then limits the invariant mass of the contributing Fock state ${\mathcal
M}^2_n >
\Lambda^2 = P^{+2} \sin^2\theta/4$ from below, so that perturbative QCD
counting rules can predict the fall-off in the jet system invariant mass
$\mathcal M$.  At large invariant mass one expects the three-quark valence
Fock state of the proton to dominate.  The segmentation of the forward
detector in azimuthal angle $\phi$ can be used to identify structure and
correlations associated with the three-quark light-cone
wavefunction.\cite{BHDP}
An interesting possibility is
that the distribution amplitude of the
$\Delta(1232)$ for $J_z = 1/2, 3/2$ is close to the asymptotic form $x_1
x_2 x_3$,  but that the proton distribution amplitude is more complex.
This ansatz can also be motivated by assuming a quark-diquark structure
of the baryon wavefunctions.  The differences in shapes of the
distribution amplitudes could explain why the $p
\to\Delta$ transition form factor appears to fall faster at large $Q^2$
than the elastic $p \to p$ and the other $p \to N^*$ transition form
factors.\cite{Stoler:1999nj} One can use also measure the dijet
structure of real and virtual photons beams
$ \gamma^* A \rightarrow {\rm Jet}_1 + {\rm Jet}_2 + A^\prime $ to
measure the shape of the light-cone wavefunction for
transversely-polarized and longitudinally-polarized virtual photons.  Such
experiments will open up a direct window on the amplitude
structure of hadrons at short distances.
The light-cone formalism is also applicable to the
description of hadrons and nuclei in terms of their nucleonic and mesonic
degrees of freedom.\cite{Miller:1999mi,Miller:2000ta}
Self-resolving diffractive jet reactions
in high energy diffractive collisions of hadrons or nuclei on
electrons or nuclei
at moderate momentum transfers can thus be used to resolve the light-cone
wavefunctions of nuclei.

high energy diffractive collisions of hadrons or nuclei on
electrons or nuclei

\section{Non-Perturbative Solutions of Light-Cone Quantized QCD}

In the discretized light-cone quantization
method (DLCQ),\cite{Pauli:1985ps}
periodic boundary conditions are introduced in
$b_\perp$ and $x^-$ so that the momenta
$k_{\perp i} = n_\perp \pi/ L_\perp$ and $x^+_i = n_i/K$ are
discrete.  A global cutoff in invariant mass of the partons in the Fock
expansion is also introduced.
Solving the quantum field theory then reduces to
the problem of diagonalizing the finite-dimensional hermitian matrix
$H_{LC}$ on a finite discrete Fock basis.  The DLCQ method has now become
a standard tool for solving both the spectrum and light-cone wavefunctions
of one-space one-time theories -- virtually any
$1+1$ quantum field theory, including ``reduced QCD" (which has both quark and
gluonic degrees of freedom) can be completely solved using
DLCQ.\cite{Dalley:1993yy,AD} Hiller, McCartor, and I
\cite{Brodsky:1998hs,Brodsky:1999xj} have recently shown that the use of
covariant Pauli-Villars regularization with DLCQ allows one to obtain the
spectrum and light-cone wavefunctions of simplified theories in physical
space-time dimensions, such as (3+1) Yukawa theory.  Dalley {\em et al.}
have also showed how one can use DLCQ with a transverse lattice to solve
gluonic $ 3+1$ QCD.\cite{Dalley:1999ii}
The spectrum obtained for gluonium states is in remarkable
agreement with lattice gauge theory results, but with a huge reduction of
numerical effort.  Hiller and I \cite{Hiller:1999cv} have shown how one
can use DLCQ to compute the electron magnetic moment in QED without
resort to perturbation theory.  One can also formulate DLCQ so that
supersymmetry is exactly preserved in the discrete approximation, thus
combining the power of DLCQ with the beauty of
supersymmetry.\cite{Lunin:1999ib,Haney:1999tk} The ``SDLCQ" method has
been applied to several interesting supersymmetric theories, to the
analysis of zero modes, vacuum degeneracy, massless states, mass gaps,
and theories in higher dimensions, and even tests of the Maldacena
conjecture.\cite{Antonuccio:1999ia} Broken supersymmetry is interesting
in DLCQ, since it may serve as a method for regulating non-Abelian
theories.\cite{Brodsky:1999xj}
The vacuum state
$\ket{0}$ of the full QCD Hamiltonian coincides with the free vacuum.
For example, as discussed by Bassetto,\cite{Bassetto:1999tm}
the computation of the spectrum
of $QCD(1+1)$ in equal time quantization requires constructing the full
spectrum of non perturbative contributions (instantons).  However,
light-cone methods with infrared regularization give the correct result
without any need for vacuum-related contributions.  The role
of instantons and such phenomena in light-cone quantized
$QCD(3+1)$ is presumably more complicated and may reside in zero
modes;\cite{Yamawaki:1998cy}
\eg, zero modes are evidently necessary to represent theories with
spontaneous symmetry breaking.\cite{Pinsky:1994yi}

Even without full non-perturbative solutions of QCD, one can envision a
program to construct the light-cone wavefunctions using measured moments
constraints from QCD sum rules, lattice gauge theory,  hard
exclusive and inclusive processes.  One is guided by theoretical
constraints from perturbation theory which dictates the asymptotic form
of the wavefunctions at large invariant mass,
$x \to 1$, and high
$k_\perp$.\cite{Lepage:1980fj,Hoyer:1990pa} One can also use constraints
from ladder relations which connect Fock states of
different particle number; perturbatively-motivated numerator spin
structures; guidance from toy models
such as ``reduced"
$QCD(1+1)$~\cite{AD}; and the correspondence to Abelian theory
for
$N_C\to 0$~\cite{Brodsky:1998jk} and the many-body
Schr\"odinger theory in the nonrelativistic domain.

\section{Conclusions}

The universal, process-independent and
frame-independent light-cone Fock-state wavefunctions
encode the properties of a hadron in terms of its fundamental quark and
gluon degrees of freedom.  Given the proton's light-cone wavefunctions,
one can compute not only the moments of the quark and gluon distributions
measured in deep inelastic lepton-proton scattering, but also the
multi-parton correlations which control the distribution of particles in
the proton fragmentation region and dynamical higher twist effects.
Light-cone wavefunctions also provide a systematic framework for
evaluating exclusive hadronic matrix elements, including time-like heavy
hadron decay amplitudes, form factors, and deeply virtual Compton
scattering.  The formalism also provides a physical factorization scheme
for separating hard and soft contributions in both exclusive and
inclusive hard processes.  A new type of jet production reaction,
``self-resolving diffractive interactions" can provide direct information
on the light-cone wavefunctions of hadrons in terms of their QCD degrees
of freedom, as well as the composition of nuclei in terms of their nucleon
and mesonic degrees of freedom.

\section*{Acknowledgments}
Work supported by the Department of Energy
under contract number DE-AC03-76SF00515.
I wish to thank Chris Pauli for his extraordinary efforts in organizing
this highly successful meeting in Heidelberg.

\end{document}